# Anisotropic and extreme magnetoresistance in the magnetic semimetal candidate Erbium monobismuthide

L. - Y. Fan[1,†], F. Tang[1,2,†], W. Z. Meng[3], W. Zhao[4], L. Zhang[1], Z. D. Han[1], B. Qian[1,*], X. -F. Jiang[1], X. M. Zhang[3,*], and Y. Fang[1,*]

[1]Jiangsu Laboratory of Advanced Functional Materials, Department of Physics, Changshu Institute of Technology, Changshu 215500, China

[2]Jiangsu Key Laboratory of Thin Films, School of Physical Science and Technology, Soochow University, Suzhou 215006, China

[3]School of Materials Science and Engineering, Hebei University of Technology, Tianjin 300130, China.

[4]ISEM, Innovation Campus, University of Wollongong, Wollongong, New South Wales 2500, Australia

[†]L. Y. Fan and F. Tang contribute equally to this work.

[*]Corresponding address: fangyong@cslg.edu.cn (Y. Fang), and njqb@cslg.edu.cn (B. Qian) zhangxiaoming87@hebut.edu.cn (X. M. Zhang)

**Abstract**

Rare-earth monopnictides display rich physical behaviors, featuring most notably spin and orbital orders in their ground state. Here, we grow ErBi single crystal and study its magnetic, thermal and electrical properties. An analysis of the magnetic entropy and magnetization indicates that the weak magnetic anisotropy in ErBi possibly derives from the mixing effect, namely the anisotropic ground state of $Er^{3+}$ ($4f^{11}$) mingles with the isotropic excited state through exchange interaction. At low temperature, an extremely large magnetoresistance ($\sim 10^4$%) with a parabolic magnetic-field dependence is observed, which can be ascribed to the nearly perfect electron-hole compensation and ultrahigh carrier mobility. When the magnetic field is rotated in the *ab* (*ac*) plane and the current flows in the *b* axis, the angular magnetoresistance in ErBi shows a twofold (fourfold) symmetry. Similar case has been observed in LaBi where the anisotropic Fermi surface dominates the low-temperature transport. Our theoretical calculation suggests that near the Fermi level ErBi shares similarity with LaBi in the electronic band structures. These findings indicate that the angular magnetoresistance of ErBi could be mainly determined by its anisotropic Fermi surface topology. Besides, contributions from several other possibilities, including the spin-dependent scattering, spin-orbit scattering, and demagnetization correlation to the angular magnetoresistance of ErBi are also discussed.

**Introduction**

Ever-increasing interest in the rare-earth-based intermetallic compounds is due, in part, to the fact that the hybridization between localized 4$f$ and conduction electrons can sometimes induce intriguing electronic and magnetic states [1-3]. A rich variety of emergent phenomena, including the heavy fermion state, complicated magnetic phases, unconventional superconductivity, valence fluctuation, non-Fermi-liquid behavior, and so on [4-8], have been frequently revealed in these compounds.

Recent attention is given to the NaCl-type rare-earth monopnictide $R$Pn ($R$=rare earth; Pn=Sb, Bi) family [9-26], which are predicted to be promising correlated topological semimetal candidates [24, 27]. $R$Pn are always identified as compensated semimetals in the electronic structure aspect, where the conduction and valence bands are constituted by the 5$d$ t$_{2g}$ state of $R$ and $^{3/2}p$ state of Pn, and lie at the X and Γ points in the first Brillouin zone, respectively [28]. The nontrivial electronic structures of $R$Pn, including the Dirac nodes or topological insulating gaps along the Γ-X line, are found to significantly depend on Pn and an unusual fourfold degenerate Dirac surface state at $\bar{M}$ [28]. The nontrivial $Z_2$-invariant and topological surface state with multiple Dirac cones are identified in LaBi that hosts empty 4$f$ shell but strong spin-orbit coupling (SOC) [20]. Once the 4$f$ shell of $R$ ion is partially filled, the magnetic transitions could take place at low temperature ( below the Néel temperature $T_N$), breaking the time-reversal symmetry and possibly yielding novel topological states with electronic correlations [24]. Strongly anisotropic magnetism and metamagnetic properties are always observed below $T_N$ in most of the magnetic $R$Pn [10,12,15,16,21,25], which significantly affects the transport behaviors. Except for the plain parabolic field dependence, the extremely large magnetoresistance (MR) also shows turning points around the magnetic transitions in $R$Pn ($R$=Ce, Nd, Dy and Ho) with Ising-like spin states [10,12,15,16,21,25]. Intriguingly, orbital order is found to partially participate in the transport properties of some members with specific spin configurations. A typical case is the type-I antiferromagnet CeSb (propagation vector $k$=[001]), where the preferred ferromagnetic cruciform $\Gamma_8^{(1)}$ orbital state of Ce ion enhances the in-plane transfer integral and thus yields high in-plane carrier mobility on the ferromagnetic plane [27]. As a consequence, with respect to other isostructural analog, lower residual resistivity but larger magnetoresistance come into being in CeSb [27]. On another hand, the orbital contribution to the transport properties of CeSb can be directly reflected from the magnetoresistance measurement [29]. In general, the angular-dependent transverse magnetoresistance of its counterparts $R$Pn ($R$=La and Gd) shows fourfold symmetric polar diagrams with the minimal values only at 0°, 90°, 180°, and 270° [30,31]. Recently, Xu *et al* found similar case in the paramagnetic phase of CeSb as well [29]. However, a series of additional minima at 45°, 135°, 225°, and 315° are observed below $T_\mathrm{N}$, which is ascribed to the enhanced electron scattering from magnetic multidomains generated by the first-order $\Gamma_8^{(1)}$ orbital-flop [29]. For comparison, the angular dependence of magnetoresistance for the orbitally quenched GdBi, which only shows a fourfold symmetric magnetoresistance as those observed in its isostructural counterparts LaSb and LaBi [30, 31], is also studied [29], clearly revealing the role of anisotropic orbital order on the magnetoresistance anisotropy in CeSb. Thus, studying structurally related compounds is a fertile ground to explore how the evolution of anisotropy effects along the series affects the ground state of their members without dealing with a more complex set of interactions competing at the same energy scale.

The physical properties of a series of magnetic $R$Pn ($R$=Nd, Sm, Dy, Ho) which are isostructural analogues of CeSb, have been intensively studied in the literature [10, 15, 16, 25]. To the best of our knowledge, scarcely few works [32-36], however, have been carried out on ErPn, which has different spin and orbital orders from CeSb [28]. Here, we report the growth of ErBi single crystal, and study the structure, magnetic susceptibility, magnetization, specific heat, resistivity, and magnetoresistance. We find that in the magnetic ground state of ErBi the first excited isotropic doublet state is admixed to the anisotropic quartet, which yields weak magnetic anisotropy and releases 72% of magnetic entropy $Rln4$ at $T_\mathrm{N}$. And for this reason, ErBi shares the similarity with LaBi and GdBi in the magnitude of residual resistivity and angular magnetoresistance. Our finding suggests that the anisotropic Fermi surface topology is largely responsible for the angular magnetoresistance of ErBi.

**Experimental**

ErBi single crystals were prepared by Bi self-flux method. Chips of Er (99.9%) and Bi powders (99.999%) were placed in alumina crucibles with an atomic ratio of 1:19 and sealed in evacuated quartz tubes. All the operations were conducted in an oxygen-free glove-box filled with nitrogen. The sealed tubes were heated up to 1473 K in 12 hours, kept for 24 hours, and then slowly cooled to 773 K in 400 hours. At this temperature, the excess Bi flux was decanted in a centrifuge. Crystal symmetry of the as-grown samples was determined using a back-reflection Laue camera. X-ray diffraction (XRD) spectrum was obtained by a Rigaku *x*-ray diffractometer. Elementary composition of the single crystals was checked by a scanning electron microscope (SEM) equipped with an energy-dispersive *x*-ray (EDX) spectrometer. Magnetization measurements were performed on a System-Vibrating Sample Magnetometer (VSM). Resistivity was measured by a conventional four-probe method and symmetrized to remove the effect of misalignment of the voltage wires ( $\rho_{xx} = \frac{\rho_{xx}(B)+\rho_{xx}(-B)}{2}$ ). Specific-heat measurements were performed using the time-relaxation method. All the physical property characterizations were conducted on a Quantum Design physical property measurement system (PPMS).

**Results and discussion**

The inset of Fig. 1(a) shows the Laue diffraction pattern for ErBi single crystal, from which the cubic symmetry can be easily determined. The powder XRD spectrum is displayed in Fig. S1 (see the Supplemental Material), where all the reflections can be indexed with a NaCl structure (space group $Fm\bar{3}m$) and no identifiable second phase is observed with the apparatus resolution, suggesting that the as-grown crystals are of good quality. Figure 1(a) plots the XRD pattern of ErBi crystal at room temperature, in which only the (*l*00) peaks can be detected, suggesting that a uniform *a*-axis is perpendicular to the plane. The EDX measurements [Fig. 1(b)] performed on several samples give an approximate atomic ratio of Er:Bi=1.08:0.92. According to the element mapping in Fig. 1(d), Er and Bi are found to be uniformly distributed across the sample surface, conforming the absence of Bi thin film or cluster.

The main panel of Fig. 2(a) exhibits the temperature-dependent specific heat $C_P$ of ErBi and the nonmagnetic reference LuBi from 2 to 50 K. As shown, at all temperatures the specific heat of ErBi is larger than that of LuBi. The sharp peak at around 3.6 K confirms the occurrence of long-rang magnetic orders, which has also been observed in the magnetic susceptibility (shown below). Besides, a broad bump arising from the thermal variation of population of the crystal-electric-field (CEF) levels is revealed, which is more pronounced in Fig. 2(b). Here, the specific heat of ErBi can be divided into three main parts: $C_{tot}=C_e+C_l+C_m$, where $C_e$, $C_l$, and $C_m$ represent the electronic, lattice and spin contributions, respectively [37]. The inset of Fig. 2(a) shows the specific heat of LuBi in temperature range from 2 to 20 K and its fit (the red line) with the expression $C_P=\gamma T+\beta T^3$ [37], from which the Sommerfeld coefficient $\gamma$ is determined as 0.3 mJ/K$^2$ mol, being of the same order of magnitude as for LaBi (0.85 mJ/K$^2$ mol) [38] and β has a value of 1.2 mJ/K$^4$ mol. Generally, the lattice contribution can be simply understood by the Debye model with the temperature-independent Debye temperature $\Theta_D$ [39]. The value of $\Theta_D$ = 171.2 K for ErBi is obtained using this relation $\Theta_D^{ErBi} = (\frac{M_{LaBi}}{M_{ErBi}})^{1/2}\Theta_D^{LaBi}$ [39], where $M$ is the molar mass and $\Theta_D^{LaBi}$ = 178 K is taken from Ref. 40. Likewise, when the above mentioned $\Theta_D^{LaBi}$ is substituted by $\Theta_D^{LuBi}$ = $(12\pi^4 NR/5\beta)^{1/3}$ = 169.2 K ($N$ = 2), $\Theta_D^{ErBi}$ takes a value of 170.9 K (close to 171.2 K). Figure 2(b) shows the magnetic contribution to specific heat, which is obtained using the usual method of subtracting the specific heat of LuBi from that of ErBi. Here, the nonmagnetic ($C_e+C_l$) part of specific heat for ErBi is assumed to approximate the total specific heat of LuBi. Obviously, the magnetic contribution extends for temperature above 30 K ($\gg T_N$), exhibiting a pronounced maximum at $T_N$ associated with bulk magnetic order due to the RKKY exchange interaction and a broad bump derived from Schottky anomalies [37]. Due to the antiferromagnetic order and broad peak associated with CEF effect, $\gamma$ for ErBi can't be simply estimated from the relation $C_P=\gamma T+\beta T^3$ [41]. Here, we fit the $C_m$ data using $\gamma'T+C_{Sch}$ (with the antiferromagnetic temperature region excluded), where $C_{Sch}$ is the Schottky specific heat [41]. As can be seen in Fig. 2(b), the reproduction of the experimental results is fairly good, from which $\Delta_1 = 6.8$ K, $\Delta_2 = 32.68$ K, and $\gamma' = 3.4$ mJ·K$^{-2}$·mol$^{-1}$ are obtained. Here, $\Delta$ is the energy separation, of which the values are close to those in early literature [42]. Besides, $\gamma = \gamma' + \gamma(LuBi) \sim 3.7$ mJ·K$^{-2}$·mol$^{-1}$ for ErBi, which is slightly larger than that of LaBi (0.85 mJ·K$^{-2}$·mol$^{-1}$) [38]. And it can also be seen that larger electronic density of states at the Fermi level $N(E_F)$ occurs in ErBi, since the electronic density of states is related to the Sommerfeld coefficient through $N(E_F) = 3\gamma/k^2_B\pi^2$ [39]. To better understand the functional dependence of the low-temperature magnetic contribution to the specific heat, we fit the data below 3.5 K with $C_m \sim AT^3e^{-E_g/T}$ which is expected to work in an antiferromagnet with an energy gap $E_g$ in the magnon dispersion relation [41]. A well fitted curve (the red line) in the inset of Fig. 2(b) yields $E_g$ with a value of approximately 0.75 K. An integral of $C_m/T$, the magnetic entropy $S_m$, is plotted in Fig. 2(c). As can be seen, the magnetic part of total entropy increases with enhanced temperature and at 50 K reaches the value of 18.3 J/K mol, which is less than the maximal value $S_m = R\ln(2J+1) = R\ln 16 = 23.05$ J/K mol. Thus, it can be concluded that thermal population of the CEF-split states in ErBi is not fully complete at this temperature. Besides, the magnetic entropy at $T_N$ is evaluated to be 8.1 J/K mol which corresponds to 1.4$R$ln2 and 0.72ln4, suggesting that the first excited state mingles with the $\Gamma_8$ quartets at CEF ground state [43].

Figure 3(a) displays the magnetic susceptibility $\chi$ as a function of temperature under an applied magnetic field of 0.01 T for ErBi in the *a* direction. As plotted, the magnetic susceptibility shows its maximal value at around 4 K. The inset of Fig. 3(a) presents the temperature-dependent inverse magnetic susceptibility $\chi^{-1}$, where the experimental data above 100 K can be well modeled by the Curie-Weiss law $\chi = C/(T-\theta_{CW})$ [37]. Here, $C$ is the Curie constant and $\theta_{CW}$ denotes the Curie-Weiss temperature. A linear fit of the $\chi^{-1}$ data gives $\theta_{CW}$ = -8.33 K, which implies the electronic correlation in ErBi is antiferromagnetic in essence [37]. Small deviation of the $\chi^{-1}$ curve from the linear fitting occurred at low temperature is possibly due to the CEF effect [44]. The effective magnetic moment $u_{eff} = \sqrt{8C}$ obtained from the paramagnetic susceptibility is estimated to be 9.4 $u_B$/Er, which is slightly less than the expected value $u_{eff} = g\sqrt{J(J+1)}u_B$=9.58 $u_B$ of a free $Er^{3+}$ ion with $g$ = 6/5 and $J$ = 15/2 [45]. This implies the presence of well-localized magnetic moments in ErBi. Figure 3(b) plots the temperature-dependent magnetic susceptibility under different fields (> 0.01 T), from which it can be found that $T_N$ shifts from 3.9 K (0.1 T) to 2.7 K (3 T), indicating that the antiferromagnetic state in ErBi can be suppressed by the external applied magnetic fields. Indeed, as shown in the top inset of Fig. 3(b), no discernible sign of antiferromagnetic transition is found above 2 K under a magnetic field larger than 4 T. Thus, some flattening of the 1/$\chi$ under larger magnetic fields can be obtained at low temperature, which corresponds to a Brillouin-like curvature of the field variation of magnetization below 10 K, as displayed in Fig. 3(c). These features make ErBi be a promising candidate for studying its magnetic-field manipulation. The bottom inset of Fig. 3(b) constructs the magnetic phase diagram for ErBi, similar to those in SmOs$_4$P$_{12}$ and CeAgAs$_2$ [37, 46], in which $T_N$ decreases with increasing magnetic field in the form of $T_N = a+b\times B^2$. Figure 3(c) exhibits the isothermal magnetization $M$ of ErBi along the $a$ direction at different temperatures between 2 K and 30 K with an interval of 1 K. The magnetization obtained at 2 K and 9 T attains a value of 6.03 $u_B$, which is significantly less than the saturation moment of Er$^{3+}$ given by $gJ$ (6/5×15/2) = 9 $u_B$ and clearly provides an indication of splitting for the ground state multiplet $^4I_{15/2}$ [45]. An extrapolation of the high-field part of the isothermal magnetization at 2 K crosses the $M$ axis at $M_0 \sim 5.62$ $u_B$. For an Er$^{3+}$ ($4f^{11}$) ion locates in an octahedral coordination, its ground-state multiplet $^4I_{15/2}$ splits into five $\Gamma_n$ levels, including two doublets ($\Gamma_6$ and $\Gamma_7$) and three quartets ($\Gamma_8^{(1)}$, $\Gamma_8^{(2)}$ and $\Gamma_8^{(3)}$), all of which carry a magnetic moment. Previous results suggest that the ground-state CEF level for ErBi is $\Gamma_8^{(3)}$ quartet [42]. Note that the four states of $\Gamma_8^{(3)}$ quartet have magnetic moments with ±6.31 $u_B$ and ±1.33 $u_B$, which seriously deviate from the experimental value [47]. Besides, the magnetic moments of $\Gamma_8^{(3)}$ quartet are aligned along the $c$ axis [42], which seems to be divergent from our results. One possible reason is that these discrepancies are caused by the mixing effect, namely, the ground state is mixed with the isotropic $\Gamma_6$ or $\Gamma_7$ excited state through the exchange interaction, which is in agreement with those obtained from the specific heat measurements.

It is clear that the isothermal magnetization in Fig. 3(c) show nonlinear field dependence even if the temperature warms up to 30 K, which suggests that the short-range of magnetic orders persists to the temperature well above $T_N$. The nonlinearity in magnetization below $T_N$ can be ascribed to the field-induced magnetic phase transitions, which has been commonly observed in $R$Pn compound with partially filled $f$ shell [10,12,15,16,21,25]. As shown in Fig. 3(d), the magnetization at 2 K increases rapidly with the field enhanced up to 3 T after which it gradually saturates. Here, two weak anomalies, which can be clearly reflected in the derivative $dM/dB$ [right axis in Fig. 3(d)], are found. The one corresponding to the low field could be attributed to the field-driven spin-flop, while the latter one may be ascribed to the antiferromagnetic to ferromagnetic phase transition induced by the applied magnetic field. For comparison, we carefully checked the magnetization and its derivative $dM/dB$ at 5 K as well in Fig. 3(d). As plotted, the magnetization initially varies linearly with increasing magnetic field and then exhibits a tendency to saturate as is expected in the paramagnetic region. Besides, $dM/dB$ at this temperature decreases gradually with the increasing magnetic field, which indicates that the anomalies in magnetization curve at 2 K are of spin origin. This is reminiscent of the isothermal magnetization in ErSb which shows similar behaviors [34], suggesting the two Er-based compounds may host the same magnetic structures.

Figure 4(a) displays the variation of resistivity ($\rho$) for ErBi as a function of temperature under various magnetic field (0 - 9 T) with an interval of 1 T. Here, the magnetic field is applied along the $a$ axis and the current flows in the $b$ axis. As shown, the zero-field resistivity monotonously decreases when the sample is cooled from 300 K down to 2 K. Note that there is an abrupt downturn around 3.8 K corresponding to $T_N$, suggesting that the spin-disordered scattering is significantly suppressed at the ordered magnetic state [14]. Under 1 T, no remarkable change emerges and the downturn below $T_N$ remains still in the $\rho$-$T$ curve. Upon the external field further increasing, the low-temperature resistivity is dramatically enhanced, exhibiting a metal-insulator-like upturn [48]. Similar behaviors have been widely reported in its other sister compounds $R$Pn, WTe$_2$, PtBi$_2$, and so on [14, 48, 49]. Usually, resistivity plateau and extreme magnetoresistance are two typical manifestation of the resistivity response to magnetic field in these materials. As shown in Fig. 4(a), the resistivity plateau seems to be absent in ErBi, which is evidenced by the minima in the temperature-dependent derivative of resistivity $\partial\rho/\partial T$ [inset on the top left of Fig. 4(a)]. However, a dip occurs at $T_m \sim 4.4$ K and 9 T, suggesting that the resistivity plateau begins to take place. It can be found that the $\partial\rho/\partial T$ curves gradually move to the right with increasing magnetic field, which indicates that lower temperature and higher field benefit the generation of resistivity plateau in ErBi. The inset on the lower right of Fig. 4(a) illustrates the variation of magnetoresistance as a function of magnetic field for ErBi at 2 K, which simply follows a $B^n$ dependence ($n = 1.96$, red solid line). Here, the obtained index $n$ ($\sim 2$) suggests that the electron densities ($n_e$) and hole densities ($n_h$) in ErBi are nearly identical [49]. Therefore, we can roughly estimate the average carrier mobility $u_{ave}$ ($u_{ave} = \sqrt{\frac{MR}{B^2}}$) from the two-band magnetoresistance model by assuming that the electron and hole densities in ErBi are perfectly compensated [25]. Fitting the magnetoresistance at 2 K yields $u_{ave}$ ~$1.27\times10^4$ cm$^2$·V$^{-1}$·s$^{-1}$, which is comparable to those of $R$Pn sister compounds [14, 25]. The compensated and high-mobility nature of the charge carriers in ErBi are further determined by the Hall resistivity measurement (see the Supplemental Material). Thus, it can be seen that the synergistic effect of carrier compensation and ultrahigh mobility may be the key ingredients for occurrence of the extreme magnetoresistance in ErBi.

Figure 4(b) shows the magnetic-field-dependent magnetoresistance for ErBi single crystal at different temperatures from 2 to 10 K. As is seen, the magnetoresistance is greatly enhanced by external magnetic field, displaying no signature of saturation with a value of $1.2\times10^4$% at 2 K and 9 T. Besides, a weak inflection in magnetoresistance emerges at around 2.7 T corresponding to the critical field for magnetic phase transition, which indicates that this anomaly is of spin origin. With the temperature warming up, the magnetoresistance drops abruptly with approximately one order of magnitude above $T_N$, possibly due to dramatic changes of the mobility and compensated nature driven by the antiferromagnetic phase transition. Above 4 K (< 10 K), the magnetoresistance gradually decreases with the increasing temperature, but always keeps its value in the same order. Further decrement from 160% to 10% in magnetoresistance is observed when the temperature increases from 50 K to 300 K, indicating that electron-phonon scattering dominates the magnetotransport at high temperature. To better understand such behavior in ErBi, we replot the field-dependent magnetoresistance at different temperatures in Kohler's law MR~$F(B/\rho_{(H=0)})$, where $\rho_{(H=0)}$ is the zero-field resistivity at a given temperature [48]. This relation follows the fact that the magnetic field enters

Boltzmann's equation in the combination ($B\tau$) where $\tau$ is the scattering time and is inversely proportional to $\rho_{(H=0)}$ [50]. For ErBi, the magnetoresistance obeys a $B^{1.96}$ dependence, which indicates that the magnetoresistance versus $(B\tau)^{1.96}$ or $(B/\rho_{(H=0)})^{1.96}$ will collapse to a simple temperature-independent curve once the numbers of carriers contributing to the transport remain still. Besides, the Kohler's rule also works when there is only a single scattering rate or several scattering rates with unchanged relative contribution [50]. As shown, the scaled data in the inset of Fig. 4(c) at 50-300 K (interval $\sim$ 50 K) fall on a single curve, while the data especially the high-field ones below 10 K in the main panel of Fig. 4(c) slightly deviates from the Kohler's rule. The breakdown of Kohler's rule in ErBi at low temperature is also widely observed in many other semimetals [25, 50], which may be derived from multiband effect or multiple scattering mechanisms. Agreement with Kohler's scaling implies that a unique temperature-dependent scattering relaxation time for the carriers and single band approximation is sufficient to explain the transport process for ErBi at high temperature.

Generally, field-induced excitonic gaps in the linear spectrum of Coulomb interacting quasiparticles are always attributed to the resistivity crossover. However, the normalized temperature-dependent magnetoresistance curves [Fig. 4(d)] at different field fall on the top of each other, suggesting that the temperature-dependent magnetoresistance remains the same for all magnetic field [48]. Therefore, the low-temperature phase is metallic rather than insulating under a higher magnetic field. To figure out the origin of up-turn behavior in ErBi, we describe the resistivity under a certain field as: $\rho(T,H) = \rho_{(H=0)}[1 + \alpha(\frac{B}{\rho_{(H=0)}})^m]$ in light of the Kohler's rule [48].

It clearly shows that the resistivity under magnetic field is composed by two parts, $\rho_0$ and $\Delta\rho = \rho_{(H=0)}\alpha(\frac{B}{\rho_{(H=0)}})^m \sim \frac{1}{\rho_{(H=0)}}$ ($m = 1.96 \sim 2$) with opposite temperature dependence, which yields a minimum or a metal-insulator-like crossover in resistivity.

As discussed in Ref. 48 for WTe$_2$, the anisotropy of Fermi surface topology can be reflected by the angular magnetoresistance. To address this issue in ErBi, we measure its magnetoresistance at a fixed temperature of 2 K by tilting the magnetic field along different crystallographic axis and keeping the current direction unchanged. The experimental setup for magnetoresistance measurement is schematically plotted in Fig. 5(a) with the current flowing along the *b* axis and the magnetic field rotating in *ab* or *ac* planes, respectively. Here, it should be noted that in the rare-earth compounds except for the Fermi surface topology, spin-dependent scattering, spin-orbit scattering and demagnetization correlation are generally considered to be the physical origin for the anisotropic magnetoresistance as well [51,52]. As compared in the Fig. 5S of Ref. 29, there is no significant difference between the angular magnetoresistance of CeSb under the anisotropic and isotropic magnetic states, indicating that the magnetic anisotropy doesn't take significant effect on the angular magnetoresistance. For ErBi, similar case can be expected, since this compound hosts weak magnetic anisotropy. Besides, as discussed below, it can be seen that the demagnetization correlation can only slightly change the shape of angular magnetoresistance. Thus, the Fermi surface anisotropy instead of the spin-dependent scattering, spin-orbit scattering and demagnetization correlation makes a major contribute to the angular magnetoresistance in ErBi. Figure 5(b) displays the variation of magnetoresistance as a function of angular at selected magnetic fields (1, 3, 5 and 7 T), from which it can be found that the magnetoresistance simply follows a $B|\cos\varphi|$ function (the red solid line), suggesting that the normal component of magnetic field governs the magnetoresistance [14]. Besides, the twofold symmetric magnetoresistance reflects the symmetry of projected profile of the Fermi surface on the plane normal to applied current [53]. Hence, the system and the Fermi surface of ErBi with an inversion and a $C_{2y}$ symmetry can be easily extracted from the angular dependence of magnetoresistance. For the current applied in the *b* axis, as shown in Fig. 5(b), the $C_{2y}$ symmetry yields $\rho_{yy}(\varphi) = \rho_{yy}(\varphi+\pi)$, $0 < \varphi < \pi$. Clearly, the magnetoresistance is highly anisotropic in Fig. 5(b) with its maximum and minimum emerging at the field orientation $\varphi = 0°$ and 90°. Therefore, the polar diagram of angular magnetoresistance for ErBi should have a peanut shape (not shown). Figure 5(c) shows the field-dependent magnetoresistance at several given angles with the field applied in the *ab* plane. As seen, the magnetoresistance decreases with the angle increasing from 0° to 90°. Note that the monotonously increasing magnetoresistance at $\varphi = 90°$ as a function of field indicates the Fermi surface of ErBi can't be two-dimensional (2D) but 3D in nature [48]. The angular magnetoresistance reminds us of this fact that the titled compound hosts an anisotropic Fermi surface topology at the ground state. As is known, the resistivity can be simply described as $\rho = \frac{m}{ne^2\tau}$ in the framework of semiclassical model, where *m*, *τ* and *e* are the effective mass, relaxation time and electron charge, respectively. It thus can be seen that the anisotropy of effective mass acts as a key ingredient in determining the anisotropic magnetoresistance and is a macroscopic reflection of the Fermi surface topology. Here, we introduce a factor $\varepsilon_\varphi$ to scale the magnetic field, and thus the magnetoresistance curves in Fig. 5(c) can collapse onto a single curve [Fig. 5(d)], namely $R(B, \varphi) = R(\varepsilon_\varphi B)$ [53]. Here, $\varepsilon_\varphi B$ is the reduced magnetic field, and $\varepsilon_\varphi = (\cos^2\varphi + \gamma^{-2}\sin^2\varphi)^{1/2}$ signifies the mass anisotropy for an elliptical Fermi surface with $\gamma^2$ being the effective mass ratios of electrons moving in the directions given by 0° and 90° [53]. This scaling operation has been widely employed to understand the angular-dependent magnetoresistance in graphite, WTe$_2$ and the anisotropic properties of high-temperature superconductors [48,54,55]. As shown in the inset of Fig. 5(d), $\gamma$ with a value of about 3.3 for ErBi is obtained by a fit to $\varepsilon_\varphi$ at 2 K as a function of angular, which is less than one-half of the corresponding value ($\gamma \sim 7.9$) obtained in LaBi [30]. Figure 5(e) shows the magnetoresistance as a function of magnetic field at several selected angular and 2 K for ErBi with the magnetic field being always perpendicular to the current. As sketched, the magnetoresistance increases once the applied magnetic field is rotated away from the *a* axis, reaches its maximum at approximately $\varphi \sim 45°$, and then gradually decreases until $\varphi \sim 90°$. The angular magnetoresistance plotted in Fig. 5(f) shows a fourfold symmetry, which is reminiscent of those in LaSb and LaBi [30, 31]. As reported, these two La-based compounds share similar Fermi surface topologies consisting of three electron pockets ($\alpha_1$, $\alpha_2$, and $\alpha_3$) and two hole pockets ($\beta$ and $\gamma$) in the first Brillouin zone [30, 31]. Note that the electron pockets in the two compounds are identical and significantly anisotropic in shape, which make crucial contributions to several distinctive electrical properties like the pseudo-2D transport phenomena, field-induced strong valley polarization, pressure-induced superconductivity, and so on [30, 31, 56, 57]. The twofold and fourfold symmetric magnetoresistance observed in LaSb and LaBi are always attributed to the anisotropy of the electron pockets in the Fermi surface as well [30, 31]. Since the magnetoresistance in ErBi shows similar angular dependence, acquisition of some clue to the detail of its electronic band structure can be expected. These findings above indicate that the Fermi surface which are analogous to those of LaBi and LaSb may emerge in ErBi. Now, one may concern the spin and orbital contributions to the electronic band structure of this compound. Indeed, in certain $R$Pn ($R$=rare earth; Pn=Sb, Bi), the electronic band structures can be significantly modified once spin order is introduced [25]. Even though, the anisotropic electron pockets always exist near the Fermi level ($E_F$). Thus, the anisotropic angular magnetoresistance could be commonly observed in this series. Such a case is verified in the antiferromagnetic GdBi with a MnO-type spin configuration, of which the ground state shows more complicated electronic band structures with respect to those in nonmagnetic members [58]. However, GdBi still exhibits the twofold and fourfold symmetric magnetoresistance, as those present in LaSb and LaBi, further verifying the significance of anisotropic electron pockets for the angular magnetoresistance. Referring to ErBi, Khalid *et al* find that in the ferromagnetic phase its electronic band structure is nearly the same as that of LaBi [59], indicating that the anisotropic electron pockets can be extracted around the $E_F$. Actually, our calculations (see the Supplemental Material) on the ferromagnetic state of ErBi confirm the presence of elongated ellipsoidal electron pockets centered at the X point in the Brillouin zone. Similar to those in other $R$Pn [14,25,50], the Hall resistivity for ErBi (see the Supplemental Material) exhibits nonlinear behavior and is negative at high field at low temperature, suggesting that the dominant carriers are electrons. Thus, it can be found that the electron pockets in the Fermi surface of ErBi also have a profound effect on its angular magnetoresistance. Note that in Fig. 5(f) the magnetoresistance at the magnetic field angle $\varphi = 90°$ and 270° are nominally smaller than those at $\varphi = 0°$ and 180°, which deviates from the results observed in LaSb and LaBi, namely the two compounds show symmetric magnetoresistance values at the same four magnetic field orientations [30,31]. This case has been observed in the angular magnetoresistance of CeSb, which is ascribed to the symmetry-breaking induced by a small tetragonal distortion or a small misalignment between the magnetic field rotation plane and the current flow direction [29]. For ErBi, the demagnetization effect could possibly result in an asymmetry in the magnetoresistance anisotropy as well. The demagnetization factor (*D*) for ErBi used in the main text are 0.38 and 0.55 (1.5×0.2×0.3 mm$^3$) for $\varphi = 0°$ and 90°, respectively. Since the internal magnetic flux density $B_{int} = B_{app} + \mu_0(1-D)M$ at $\varphi = 90°$ is slightly smaller than that at $\varphi = 0°$, a tiny difference of the magnetoresistance between these two $\varphi$s can come into being. Here, $B_{app}$ is the applied magnetic field. Nevertheless, it is clear that the shape of angular magnetoresistance of ErBi is quite similar to that of LaBi, suggesting that Fermi surface topology plays a major role in determining the magnetoresistance anisotropy in ErBi.

**Conclusion**

  We have successfully grown the cubic ErBi single crystals using Bi-self flux method, and systematically studied their magnetic, electrical, and thermal properties by means of susceptibility, magnetization, resistivity, magnetoresistance, specific heat and so on.

Our data reveals that ErBi orders antiferromagnetically at $T_N$ ~ 4 K which decreases with increasing magnetic field. Below $T_N$, field-induced phase transitions are observed in the isothermal magnetization curves. Resistivity measurement evidences that ErBi is metallic at the ground state and a given magnetic field can yield upturn behavior. Normalized magnetoresistance at different field clarifies that the metal-insulator-like transition in resistivity does not originate from gap opening effect. Besides, anisotropic magnetoresistance with twofold and fourfold symmetries are also observed in ErBi, which resemble those of LaSb and LaBi. Our electronic band structure calculations suggest that ErBi is an antiferromagnetic semimetal candidate with the same Fermi surface topology as obtained in the La-based sister compounds. Besides, we find that the angular magnetoresistance is barely affected by the spin and orbital orders, but mainly dominated by the anisotropy of Fermi surface morphology.

**Acknowledgments**

This work is supported by the National Natural Science Foundation of China (Grant No. 11604027 and U1832147), Key University Science Research Project of Jiangsu Province of China (19KJA530003), Open Fund of Fujian Provincial Key Laboratory of Quantum Manipulation and New Energy Materials (Grant No. QMNEM1903).

**Figure caption**

FIG. 1 (a) XRD patterns for ErBi single crystal with the x-ray along the perpendicular direction of the cubic surfaces. Inset shows the x-ray Laue backscattering pattern with a fourfold rotation symmetry. (b) EDX spectrum of ErBi single crystal. (c)-(d) Elemental distributions for Er and Bi, respectively.

FIG. 2 (a) Specific heat of ErBi and LuBi as a function of temperature. The inset plots the low-temperature part of $C_p$ versus $T$ of LuBi. (b) Temperature-dependent magnetic contribution $C_m$ and its fit (red line in the inset) below 3.4 K. (c) Variation of magnetic entropy for ErBi as a function of temperature.

FIG. 3 (a) Temperature dependence of the magnetic susceptibility measured at 0.01 T along the $a$ axis. Inset shows the inverse magnetic susceptibility as a function of temperature. (b) Temperature-dependent magnetic susceptibility under 0.1, 1, 2, and 3

T. Inset shows the magnetic susceptibility versus temperature under 4, 5, 6, 7, 8, and 9 T. (c) Isothermal magnetization at various temperatures ranging from 2 to 30 K with an interval of 1 K. (d) Field-dependent magnetization and its derivative at 2 and 5 K.

FIG. 4 (a) Temperature-dependent resistivity at different magnetic field. Upper inset shows the derivative $\partial\rho/\partial T$ at 3, 6, and 9 T. Lower inset plots the magnetoresistance at 2 K and its fit. (b) Magnetoresistance as a function of field at 2-10 K (interval ~2 K). Inset illustrates the magnetic-field-dependent magnetoresistance at 50-300 K (interval ~50 K). (c) Kohler's scaling of the magnetoresistance at 2-10 K and 50-300 K (inset). (d) Normalized magnetoresistance and magnetoresistance (inset) as a function of temperature at different magnetic field.

FIG. 5 (a) Schematic diagram for angular-dependent magnetoresistance measurement with field rotating in *ab* and *ac* plane. $\varphi$ is the angle between the field orientation and the *a* axis. (b) Angular-dependent magnetoresistance under 1, 3, 5, and 7 T with the magnetic field rotating in *ab* plane. (c) Magnetoresistance versus magnetic field at different angles in *ab* plane. (d) Magnetoresistance plotted as a function of $\varepsilon_0 B$, where $\varepsilon_0$ is a scaling factor $\varepsilon_\varphi = (\cos^2\varphi + \gamma^{-2}\sin^2\varphi)^{1/2}$. (e) Field-dependent magnetoresistance at different angles in *ac* plane. (f) Magnetoresistance as a function of angular at 1, 3, 5, and 7 T with the magnetic field rotating in *ac* plane.

**Figure 1**

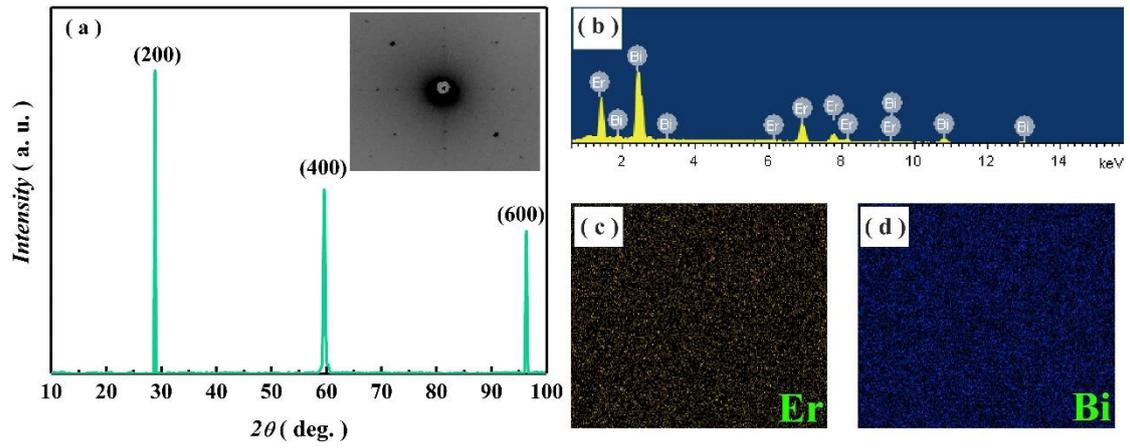

Figure 2

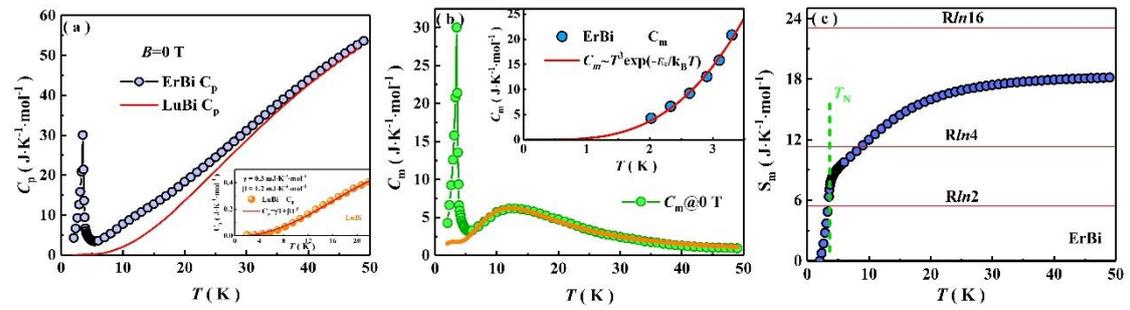

**Figure 3**

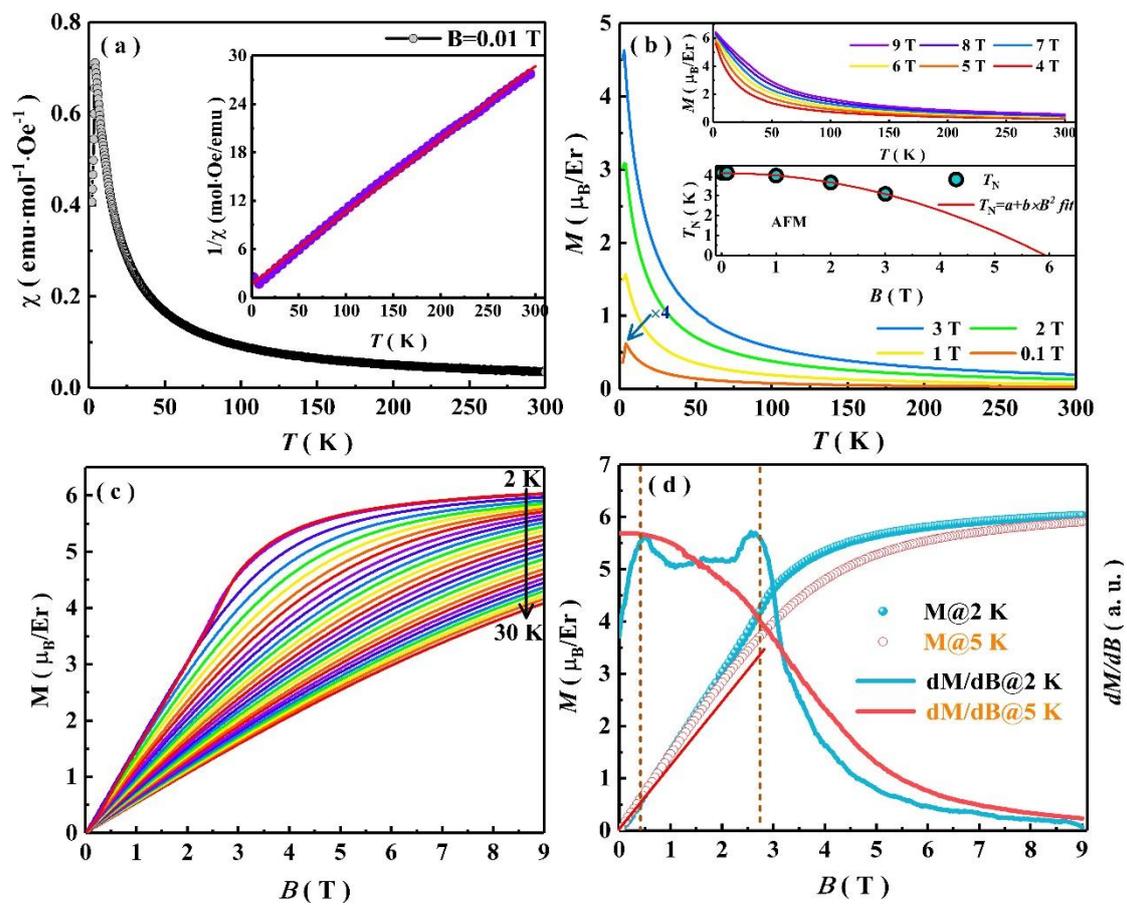

**Figure 4**

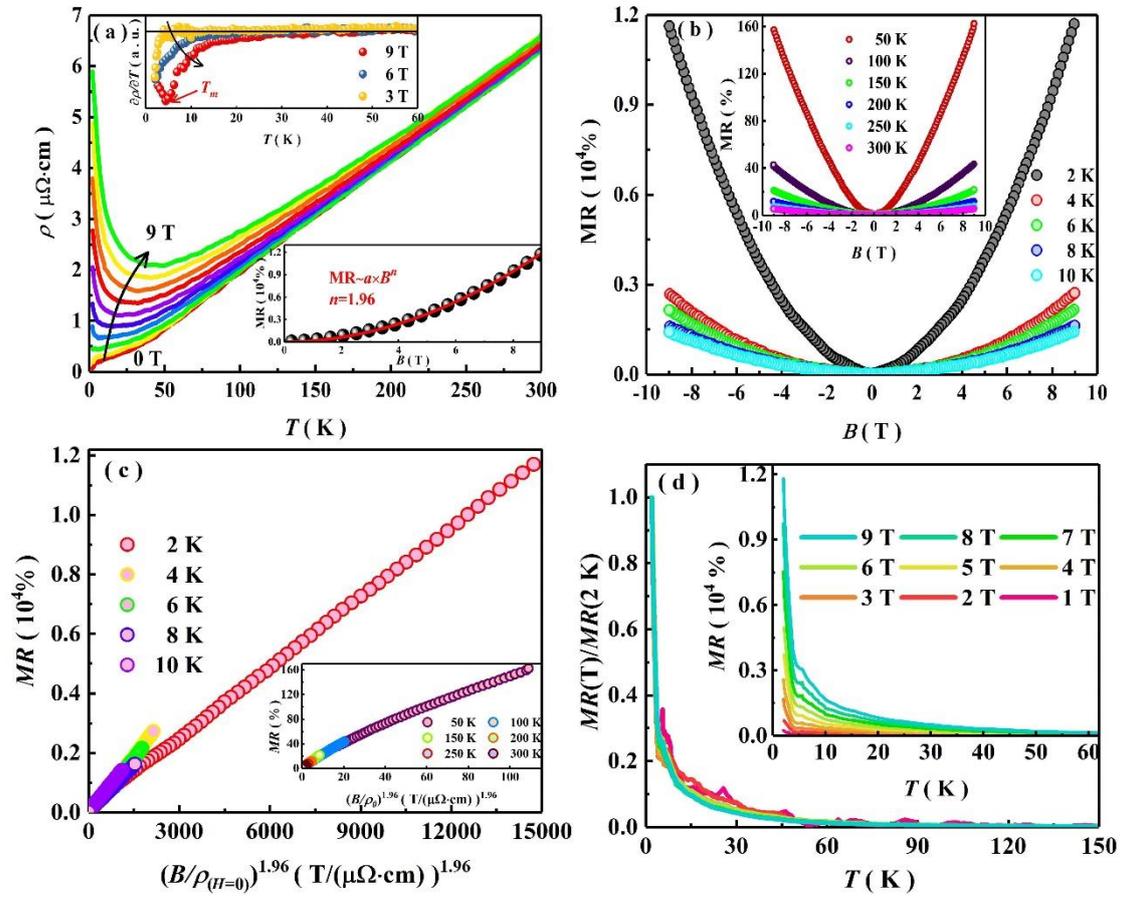

**Figure 5**

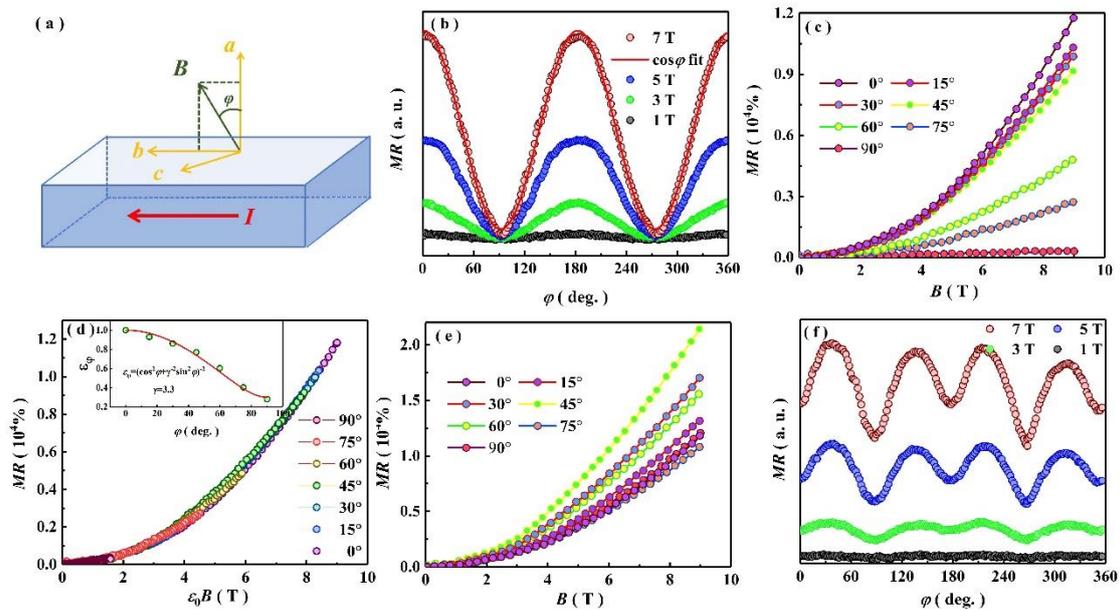